\documentclass{article}

\usepackage{microtype}
\usepackage{graphicx}
\usepackage{subfigure}
\usepackage{booktabs} 

\usepackage{hyperref}
\usepackage{multicol}
\usepackage{multirow}
\usepackage{enumitem}

\usepackage[accepted]{icml2025}

\usepackage{amsmath}
\usepackage{amssymb}
\usepackage{mathtools}
\usepackage{amsthm}
\usepackage{graphicx}
\usepackage{subcaption}

\usepackage[capitalize,noabbrev]{cleveref}

\theoremstyle{plain}

\theoremstyle{definition}

\theoremstyle{remark}

\newcommand{\TROVE}{\textsc{Tro}\textsc{V}\textsc{E}}
\newcommand{\PRIMITIVE}{\textsc{Primitive}}
\newcommand{\CODELLAMA}{\textsc{Code}\textsc{L}\textsc{La}\textsc{Ma}}
\newcommand{\CODELLAMATWO}{\textsc{Code}\textsc{L}\textsc{La}\textsc{Ma}\textsc{2-}\textsc{7b-}\textsc{Instruct}}
\newcommand{\SKIP}{\textsc{Skip}}
\newcommand{\IMPORT}{\textsc{Import}}
\newcommand{\CREATE}{\textsc{Create}}

\usepackage[textsize=tiny]{todonotes}
\usepackage{booktabs, colortbl}

\usepackage[noend]{algpseudocode}   

\usepackage{xcolor}
\algrenewcommand{\algorithmiccomment}[1]{%
  \hfill\textcolor{gray}{//\,#1}}

\icmltitlerunning{A Compute-Matched Re-Evaluation of \textsc{TroVE} on \textsc{MATH}}

\begin{document}

\twocolumn[
\icmltitle{A Compute-Matched Re-Evaluation of \textsc{TroVE} on \textsc{MATH}}



\icmlsetsymbol{equal}{*}

\begin{icmlauthorlist}
\icmlauthor{Tobias Sesterhenn}{xxx}
\icmlauthor{Ian Berlot-Attwell}{yyy}
\icmlauthor{Janis Zenkner}{xxx}
\icmlauthor{Christian Bartelt}{xxx}
\end{icmlauthorlist}

\icmlaffiliation{xxx}{Clausthal University of Technology, Clausthal, Germany}
\icmlaffiliation{yyy}{Vector Institute, University of Toronto, Toronto, Canada}

\icmlcorrespondingauthor{Tobias Sesterhenn}{tobias.sesterhenn@tu-clausthal.de}

\icmlkeywords{Machine Learning, ICML}

\vskip 0.3in]



\printAffiliationsAndNotice{}  


\begin{abstract}
Reusing established theorems and formulas is central to mathematical problem solving, serving as essential building blocks for tackling increasingly complex challenges.
Recent work, \textsc{TroVE}, argues that code-generating Large Language Models (LLMs) can benefit similarly on the \textsc{MATH} benchmark by inducing and reusing higher-level toolboxes. By allocating computational budget across an ensemble of three modes -- directly generating code, creating tools, and reusing tools -- \textsc{TroVE} claims to outperform a \PRIMITIVE{} baseline that only performs direct generation. However, recent analysis~\cite{berlot2024library} casts doubt on these gains, noting that the tools created are often trivial or rarely reused, suggesting that improvements may stem from self-consistency or self-correction.
In this work, we re-evaluate \TROVE{} on \textsc{MATH}, analyze the impact of each of its modes, and show that its benefit does not come from these mechanisms, but simply from a higher computational budget spent for \TROVE{} compared to \PRIMITIVE{}. 
To this end, we also perform a small correction in the original implementation of \TROVE{}’s selection mechanism, boosting \TROVE{}’s performance on \textsc{MATH} by $3\%$ in accuracy.
After matching for compute, the benefit of \TROVE{} reduces to a marginal improvement of $1\%$, suggesting that this toolbox approach does not provide a significant benefit on \textsc{MATH}.




\end{abstract}

\section{Introduction}

Reusing proven lemmas and formulas is how mathematicians keep cognitive load manageable and proofs short: recalling a schema (e.g., the Binomial theorem) frees working memory that would otherwise be spent on re-derivation~\citep{sweller1988cognitive,zhao2024model}, and automated–reasoning systems that reuse earlier proofs solve new theorems markedly faster than systems that start from scratch~\citep{walther2000proving}. 
In other words, good mathematics relies on a library of lemmas that can be imported rather than rebuilt.

This reuse principle has long guided library learning in computer science.  Systems such as DreamCoder~\citep{ellis2023dreamcoder} interleave solving tasks with compressing recurring code snippets into a growing library; the learned abstractions then make subsequent synthesis significantly faster and more sample-efficient.  
Later variants \citep{bowers2023top,cao2023babble,grand2023lilo} refine the idea, but the core insight remains: discover reusable components, store them, and apply them later.  
Recent work has explored how to leverage In-Context Learning (ICL), where a LLM is prompted with a description of the domain, examples, a set of available functions, and the task to solve. This allows for an incrementally extended library, or ``toolbox``, through a mechanism to select valid abstractions without retraining the model.
These approaches~\citep{zhiruo2025inducing, wang2024trove,cai2023large} have recently gained attention for their effectiveness in improving performance on various benchmarks.

\begin{table*}[t]
\centering
\definecolor{gain}{RGB}{220,255,220}
\caption{Accuracy comparison across MATH categories. We report mean and standard variation on $5$ different random seeds. The left two columns show the improvement in performance when matching \PRIMITIVE{} for compute. On the right we reproduce the results of \TROVE{} and show the performance gain when changing \TROVE{}'s selection mechanism from a two-stage to a one-stage approach. Green highlights represent strictly higher improvements for the compute-matched version of \PRIMITIVE{} and the corrected version of \TROVE{}.}
\label{tab:repro-results}
\begin{tabular}{lcc!{\vrule width 1pt}cc|c}
\toprule
& \multicolumn{2}{c!{\vrule width 1pt}}{\textbf{PRIMITIVE: Reproduction}} & \multicolumn{3}{c}{\textbf{TroVE: Reproduction}}  \\
Category & Original Results & Compute-Matched & Original Results & Reproduced & Improved \\
\midrule
Algebra & 0.15 & \cellcolor{gain}\textbf{0.27 $\pm$ 0.01} & 0.25 & 0.26 $\pm$ 0.01 & \cellcolor{gain}\textbf{0.29 $\pm$ 0.01} \\
Counting & 0.14 & \cellcolor{gain}\textbf{0.24 $\pm$ 0.00} & 0.26 & 0.24 $\pm$ 0.02 & \cellcolor{gain}\textbf{0.27 $\pm$ 0.02} \\
Geometry & 0.06 & \cellcolor{gain}\textbf{0.08 $\pm$ 0.01} & 0.08 & 0.05 $\pm$ 0.01 & \textbf{0.08 $\pm$ 0.01} \\
Intermediate & 0.05 & \cellcolor{gain}\textbf{0.14 $\pm$ 0.01} & 0.11 & 0.12 $\pm$ 0.02 & \cellcolor{gain}\textbf{0.13 $\pm$ 0.01} \\
Number & 0.16 & \cellcolor{gain}\textbf{0.28 $\pm$ 0.02} & 0.25 & 0.27 $\pm$ 0.01 & \cellcolor{gain}\textbf{0.30 $\pm$ 0.01} \\
Prealgebra & 0.21 & \cellcolor{gain}\textbf{0.33 $\pm$ 0.02} & 0.29 & 0.29 $\pm$ 0.02 & \cellcolor{gain}\textbf{0.32 $\pm$ 0.02} \\
Precalculus & 0.10 & \cellcolor{gain}\textbf{0.15 $\pm$ 0.01} & 0.17 & 0.18 $\pm$ 0.03 & \cellcolor{gain}\textbf{0.20 $\pm$ 0.01} \\
\midrule
MATH & 0.12 & \cellcolor{gain}\textbf{0.24 $\pm$ 0.01} & 0.20 & 0.22 $\pm$ 0.01 & \cellcolor{gain}\textbf{0.25 $\pm$ 0.01} \\
\bottomrule
\end{tabular}
\end{table*}

Among them, \TROVE{}~\citep{wang2024trove} currently stands out as a state-of-the-art method on the MATH dataset~\cite{hendrycks2021measuring}. 
For a fixed computational budget of $K$ LLM calls, \TROVE{} repeatedly generates Python code to solve a given task using three different prompting modes and then selects the most likely correct solution via self-consistency (i.e., majority vote)~\citep{wang2022self}. The three modes are:
\begin{itemize}
	\item \textbf{\CREATE{}:} Solve the task and add any new functions to a toolbox that is periodically pruned.
	\item \textbf{\IMPORT{}:} Solve the task by using functions already present in the toolbox.
	\item \textbf{\SKIP{}:} Solve the task only with primitive, built-in functions, without using the toolbox.
\end{itemize}
Hence, when evaluating the use of \CODELLAMA{}~\citep{roziere2023code}, \citet{wang2024trove} reported that \TROVE{}  outperforms a \PRIMITIVE{} baseline that operates only in \textsc{SKIP} mode achieving $20\%$ on MATH compared to \PRIMITIVE{} only solving $12\%$.

However, \citet{berlot2024library} raise the question whether the \TROVE{} mechanism actually increases performance as they show that tools are either trivial (already contained in the knowledge of the LLM) or never reused (see Appendix \ref{apx:additional_experiments}):
An ablation study in a subset of MATH shows that the removal of the \IMPORT{} mode does not negatively affect performance.
Although the authors hypothesize \TROVE{}'s benefit may come from self-consistency or the self-correctness mechanism, the question of what makes \TROVE{} better remains an open question.

In this work, we investigate what really drives \TROVE{}’s strong performance on the MATH benchmark and show that its main advantage comes from a higher computational budget rather than the toolbox mechanism itself.
We first test whether \TROVE{} continues to outperform a \PRIMITIVE{} baseline when both systems are given the same number of LLM calls.
To ensure a fair comparison, we also fix a discrepancy between \TROVE{}’s described agreement-based selection method and its actual implementation, correcting an error that improves its performance by $3\%$ accuracy.
Finally, we analyze whether \TROVE{}’s varied prompting strategies provide benefits independently of the toolbox, and how \TROVE{} and \PRIMITIVE{} scale with increased compute under an ideal selection mechanism.

Our results show that while \TROVE{} marginally outperforms \PRIMITIVE{}, the accuracy gain reduces to only $1\%$ when controlling for the computational budget. 
We also observe that, under compute-matched settings, \TROVE{} produces a slightly more diverse set of candidate predictions --- on average 0.74 more per task. 
While this may offer some benefit by broadening the search space, it also increases the difficulty in selecting the correct answer.
Our findings indicate that \TROVE{}’s improved performance is best explained by a higher allocation of the computational budget, rather than by its toolbox mechanism.


\section{Revisiting \TROVE{}}
\noindent{\textbf{Prompting Modes}} \TROVE{} is a training-free method to build a toolbox of reusable high-level functions to solve programmatic tasks, operating without ground truth labels or external supervision~\citep{wang2024trove}. 
For each task, \TROVE{} samples $K$ candidate programs using a fixed computational budget, divided equally among three distinct prompting modes: \SKIP{}, \CREATE{}, and \IMPORT{}. 
In \SKIP{} mode, the model generates a solution using only primitive Python functions that, for MATH, are inherently stored in the LLM weights, i.e., parametric knowledge.
For example, the LLM can call \textit{numpy} or \textit{sympy} functions.
In \CREATE{} mode, the model is instructed to define a new helper function that encapsulates reusable logic and adds it to the toolbox. 
This function is designed to support solving the current task and potentially benefit future tasks. 
In \IMPORT{} mode, the model can select and apply existing functions from the current toolbox to solve the task.
The original work compares \TROVE{} against a \PRIMITIVE{} baseline.
Hereby, the baseline implements the same prompting mode as \SKIP{}.
Thus, when fairly comparing \PRIMITIVE{} and \TROVE{}, each TROVE mode is used $\frac{K}{3}$ times, whereas \PRIMITIVE{} is used $K$ times. 

\noindent{\textbf{Agreement-Based Selection}} To decide on a final candidate, the authors propose an agreement-based selection algorithm using majority voting and solution complexity: among the candidates that execute successfully (i.e., without errors), the most frequent answer is chosen. If there is a tie, the solution implemented by the shortest program, in terms of operations, is selected (see Algorithm \ref{alg:corrected}).

\section{Experimental Setup}
\noindent{\textbf{Compute-Matching}}
In this work, we analyze whether \TROVE{}'s methodology using library learning and three different prompting styles brings an advantage over a \PRIMITIVE{} baseline.
For this we reproduce the results of the original work by \citet{wang2024trove} and match \PRIMITIVE{} for a computational budget of $K=15$ LLM calls.
In line with the original work, we evaluate the approaches using \CODELLAMATWO{}~\citep{roziere2023code}.
We choose the hyperparameters accordingly, as described in Appendix \ref{apx:exp_setup}.
All experiments are run five times, and we report the mean and standard variation of the accuracies in each category of MATH.
Here, a task is correctly solved if the selection mechanism selects a candidate whose execution result agrees with the task solution.



\noindent{\textbf{Oracle Selection Mechanism}}
In the experiments, we also focus on the potential of \TROVE{} under an \textit{oracle selection mechanism}, which assumes a perfect selector that identifies the correct answer as soon as it appears in the candidate set. We report the corresponding performance using \textit{pass@K} for a budget of $K$.

\noindent{\textbf{MATH Dataset}}
The \textsc{MATH} dataset introduced by \citet{hendrycks2021measuring} consists of $12,500$ competition-style math problems.
In this work, we use a subset of $3,201$ tasks, following \citet{wang2024trove}.
The dataset is divided into seven categories:
$881$ tasks in \textit{prealgebra}, $291$ tasks in \textit{algebra}, $237$ in \textit{number theory}, $503$ in \textit{counting \& probability}, $497$ in \textit{geometry}, $636$ in \textit{intermediate algebra}, and $156$ tasks in \textit{precalculus}. 
Each task consists of a query in natural language and a numeric ground-truth answer.

\section{Analysis}
We first reproduce the experiments of the original work, but matching the computational budget (\S\ref{sec:align_compute}) and correcting a discrepancy in the agreement-based selection mechanism (\S\ref{sec:correcting_selection}).
We further investigate the impact of the different modes on the overall performance of \TROVE{} (\S\ref{sec:diverse_prompting}) and analyze the performance of \PRIMITIVE{} and \TROVE{} under different computational budgets (\S\ref{sec:comp_budgets}).

\subsection{Correcting Original \TROVE{} Results}
\subsubsection{Matching Computational Budget} \label{sec:align_compute}
To strengthen the hypothesis that \TROVE{}'s toolbox mechanism has no significant impact on performance, we compare \TROVE{} against the compute-matched \PRIMITIVE{} baseline across all MATH domains by sampling the LLM with $K=15$ times per task.
We further reproduce the reported results of the \TROVE{} approach, achieving $\pm 3\%$ of the original reported results over five random seeds.

As depicted in Table~\ref{tab:repro-results}, the compute-matched \PRIMITIVE{} baseline now, instead of reported $12\%$, achieves $24\%$ in MATH, outperforming the original 22\% reported by \TROVE{}.
Among the different categories, the performance differs at most $\pm 4\%$ in accuracy between \PRIMITIVE{} and \TROVE{} with \PRIMITIVE{} performing better in algebra, geometry, intermediate algebra, number and prealgebra.
When analyzing the performance across different computational budgets, the results suggest that the reported results of \PRIMITIVE{} in the original work were obtained by setting $K$ to a much smaller value than $15$ (see Appendix \ref{apx:reproduction}, Table \ref{tab:sampling_orig}).
Thus, contrary to the claim of the authors, \TROVE{} appears to benefit primarily from repeated sampling rather than toolbox learning.
Preliminary results (see Appendix~\ref{apx:domains}) suggest that this observation also applies to domains beyond MATH, such as TabMWP~\citep{lu2022dynamic}, WTQ~\citep{pasupat2015compositional}, and HiTab~\citep{cheng2021hitab}.

\subsubsection{Correcting \TROVE{}'s selection mechanism}
\label{sec:correcting_selection}

\begin{algorithm}[H]
\small
\caption{Two-Stage Candidate Selection}
\begin{algorithmic}
\Function{MultiWayGeneration}{example, K}
  \State $candidates \gets []$ 
  \ForAll{mode \textbf{in} \{IMPORT, CREATE, SKIP\}} 
    \State $c \gets \Call{SampleModel}{example, mode, K}$
    \State $i^\ast \gets \Call{SelectBest}{c}$ 
    \State \Call{Append}{$candidates$,\; $c[i^\ast]$}
  \EndFor
  \State $j^\ast \gets \Call{SelectBest}{candidates}$ 
  \State \Return $candidates$[$j^\ast$].mode,\; $candidates$[$j^\ast$]
\EndFunction
\end{algorithmic}
\label{alg:old}
\end{algorithm}

\begin{algorithm}[H]
\small
\caption{One-Stage Candidate Selection}
\label{alg:corrected}
\begin{algorithmic}[1]
\Function{MultiWayGeneration}{example, K}
  \State $candidates \gets [\,]$                       
  \ForAll{mode \textbf{in} \{IMPORT, CREATE, SKIP\}}
    \State $c \gets  \Call{SampleModel}{example, mode, K}$
    \State \Call{Extend}{$candidates$, $c$}
  \EndFor
  \State $i^\ast \gets \Call{SelectBest}{candidates}$  
  \State \Return $candidates[i^\ast].\text{mode}$,\; $candidates[i^\ast]$
\EndFunction
\end{algorithmic}
\end{algorithm}

When reviewing \TROVE{}, we also identified a difference between the algorithm described in the original work and its implementation in Python\footnote{\url{https://github.com/zorazrw/trove/blob/main/run_trove.py}}.
The original implementation proposes a single stage agreement-based selection mechanism, based on self-consistency and solution complexity on the $K$ candidate responses \cite{wang2024trove}.
However, the implementation performs a two-stage candidate selection, where in the first stage for each of the three modes, one of $\frac{K}{3}$ candidates is chosen individually and then the algorithm is applied to the three remaining candidates (see Algorithm \ref{alg:old}).

However, this leads to a lower probability of selecting a correct answer, as the candidate is chosen by majority voting on a much smaller size.
Once we correct the implementation for a one-stage candidate selection, as described in Algorithm \ref{alg:corrected}, \TROVE{}'s performance is consistently improved across all categories.
Table~\ref{tab:repro-results} shows that compared to our reproduced results, the performance increases by $3\%$ on MATH.
Compared to the originally reported results, it increases by $5\%$.
Through this, \TROVE{} slightly outperforms \PRIMITIVE{} by $1\%$, although these results are not significant (we show our findings with $5$ random seeds).

\subsection{Effects of \TROVE{}'s diverse prompting}
\label{sec:diverse_prompting}

After matching the compute, \PRIMITIVE{} performs almost equally to \TROVE{} with \TROVE{} performing only slightly better after changing to the one-stage selection mechanism.
However, \TROVE{}'s three-mode prompting mechanism brings more diversity into the approach, as the \PRIMITIVE{} prompting mode only matches the \SKIP{} mode. 
Therefore, in the following, we further analyze the impact of each prompt mode on the overall performance and compare \TROVE{}'s candidate proposals with the ones by \PRIMITIVE{}. 

To examine the potential of both \PRIMITIVE{} and \TROVE{}, in the following we use the oracle selection mechanism for candidate selection and report \textit{pass@K}.

\subsubsection{Performance of Prompt Modes}
\begin{figure}[h!]
    \centering
    \includegraphics[width=\linewidth]{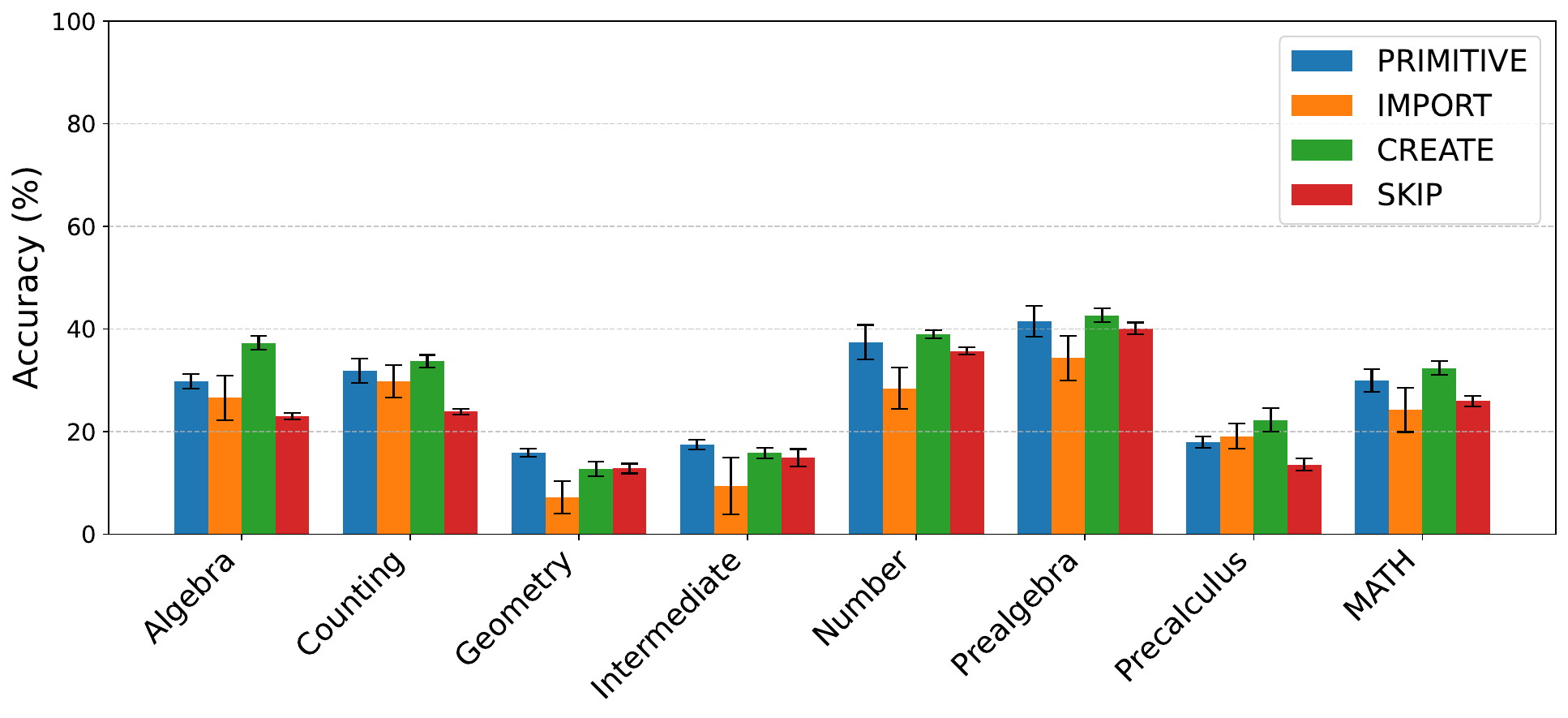}
    \caption{Distribution of solved tasks across the different categories for each mode. We report $pass@5$, i.e., using the oracle selection mechanism for a computational budget of $5$.}
    \label{fig:solved_across_modes}
\end{figure}

Figure \ref{fig:solved_across_modes} shows the distribution of the tasks solved in the different modes for $pass@5$ and across $5$ seeds.
Thereby, the \CREATE{} mode performs best across all categories, followed by \SKIP{} and \IMPORT{}.
However, \TROVE{} profits from all three modes, as they partially solve different tasks on the MATH dataset (Table~\ref{tab:unique_solutions_frac}).
While \CREATE{} uniquely solves the highest proportion of tasks, namely $8\%$ in the entire dataset, \SKIP{} and \IMPORT{} uniquely solve $6\%$ and $4\%$, respectively.

Furthermore, in Appendix \ref{apx:jaccard} we show that \TROVE{} also solves tasks that are not solved by \PRIMITIVE{} (Table~\ref{tab:trove-benefit}) and that the solutions found by \PRIMITIVE{} have the highest overlap with those found by the \SKIP{} mode of \TROVE{} (Figure~\ref{fig:jaccard_across_modes}). 
This is reasonable since both \PRIMITIVE{} and \SKIP{} use the same prompt.
In accordance with this, \PRIMITIVE{} is least similar to \textsc{IMPORT} in their sets of solved challenges, since their prompting styles differ the most.

\begin{table}[ht]
\centering
\caption{Fraction of tasks uniquely solved by each mode (mean $\pm$ standard variation over seeds). The MATH row aggregates across all categories.}
\label{tab:unique_solutions_frac}
\small
\begin{tabular}{lccc}
\toprule
Category & \textsc{IMPORT} & \textsc{CREATE} & \textsc{SKIP} \\
\midrule
Algebra & 0.04 $\pm$ 0.01 & \textbf{0.10 $\pm$ 0.02} & \underline{0.04 $\pm$ 0.01} \\
Counting & 0.05 $\pm$ 0.01 & \textbf{0.07 $\pm$ 0.01} & \underline{0.05 $\pm$ 0.01} \\
Geometry & 0.03 $\pm$ 0.01 & \underline{0.07 $\pm$ 0.02} & \textbf{0.07 $\pm$ 0.01} \\
Intermediate & 0.03 $\pm$ 0.02 & \underline{0.07 $\pm$ 0.02} & \textbf{0.08 $\pm$ 0.01} \\
Number & 0.05 $\pm$ 0.01 & \textbf{0.08 $\pm$ 0.01} & \underline{0.06 $\pm$ 0.01} \\
Prealgebra & 0.05 $\pm$ 0.01 & \textbf{0.07 $\pm$ 0.01} & \underline{0.07 $\pm$ 0.00} \\
Precalculus & \underline{0.04 $\pm$ 0.02} & \textbf{0.06 $\pm$ 0.02} & 0.03 $\pm$ 0.01 \\
\midrule
MATH & 0.04 $\pm$ 0.00 & \textbf{0.08 $\pm$ 0.01} & \underline{0.06 $\pm$ 0.00} \\
\bottomrule
\end{tabular}
\end{table}

\begin{table}[h!]
\caption{Mean number and standard variation of distinct solutions per task under a fixed compute budget of $15$ LLM calls. Positive $\Delta$ values indicate that \TROVE{} proposes a higher variety of predictions compared to the \PRIMITIVE{} baseline.}
\label{tab:solution_suggestions}
\centering
\begin{tabular}{lcc|c}
\toprule
Category & \PRIMITIVE{} & \TROVE{} & $\Delta$ \\
\midrule
Algebra & $4.20 \pm 0.04$ & $4.83 \pm 0.15$ & +0.64 \\
Counting & $5.94 \pm 0.06$ & $6.77 \pm 0.13$ & +0.83 \\
Geometry & $6.79 \pm 0.66$ & $6.64 \pm 0.41$ & -0.15 \\
Intermediate & $3.89 \pm 0.06$ & $4.09 \pm 0.25$ & +0.20 \\
Number & $4.88 \pm 0.64$ & $5.78 \pm 0.19$ & +0.90 \\
Prealgebra & $5.18 \pm 0.04$ & $6.47 \pm 0.16$ & +1.29 \\
Precalculus & $3.36 \pm 0.12$ & $4.83 \pm 0.14$ & +1.47 \\
\midrule
MATH & $4.76 \pm 0.19$ & $5.50 \pm 0.19$ & +0.74 \\
\bottomrule
\end{tabular}
\end{table}

\subsubsection{Prompt diversity increases hypothesis space per task}
Furthermore, prompt diversity actually leads to a higher variety of proposed solutions per task, suggesting that \TROVE{} is capable of covering a larger hypothesis space.
We quantify this by counting the mean number of different predictions produced per task in the fixed compute budget of $K=15$ LLM calls.  
Table \ref{tab:solution_suggestions} shows that on counting, number, prealgebra, and precalculus, \TROVE{} creates approximately one additional solution candidate, with, on average, $0.74$ additional solutions on any MATH task.
However, while the increased number of candidate solutions may be beneficial when combined with a good selection mechanism, this may not hold for a weaker one such as majority voting, where the extra predictions inject noise.

\subsection{Effect on different computational budgets}
\label{sec:comp_budgets}
\begin{figure}[!ht]
    \centering
    \includegraphics[width=\linewidth]{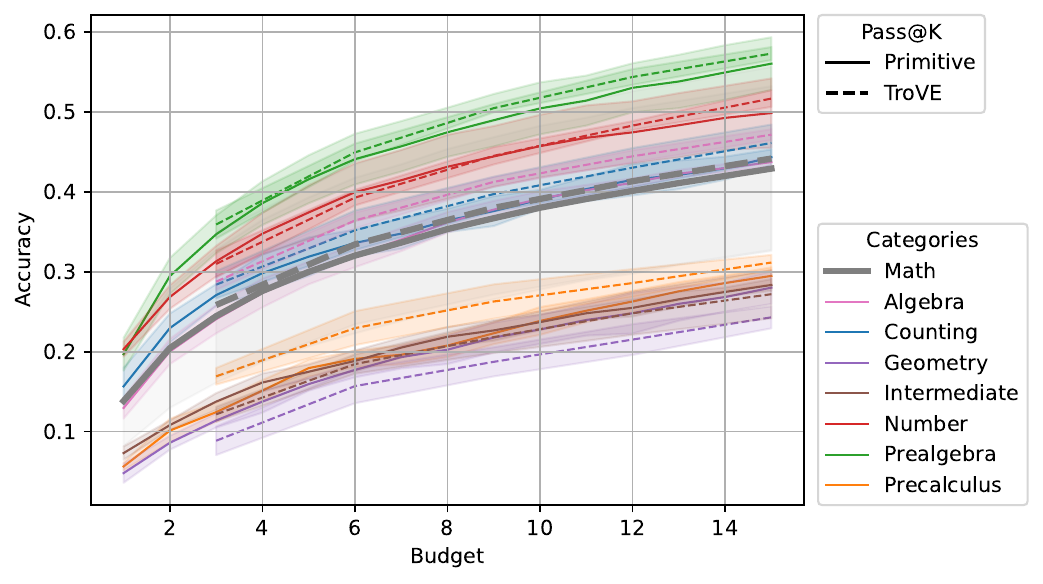}
    \caption{Performance across different computational budgets represented by the sampling sizes for \TROVE{} and \PRIMITIVE{} on the MATH dataset. As \TROVE{} requires each mode to be called, the plot depicts the performance only for multiples of $3$.}
    \label{fig:sampling_oracle_prim_trove}
\end{figure}

To analyze whether \TROVE{}'s diverse prompting leads to better pass@K results across growing computational budgets, we calculate the accuracies over different computational budgets.
However, Figure~\ref{fig:sampling_oracle_prim_trove} shows that there is no significant difference between \TROVE{} and \PRIMITIVE{}.
Whereas \TROVE{} solves $44\%$ of tasks with a budget of $K=15$, \PRIMITIVE{} achieves $43\%$ accuracy.
Figure~\ref{fig:n_75_accuracies} in Appendix~\ref{apx:selection_mechanism} shows that this gap remains consistent for a budget of up to $K=75$.
Notably, the results also highlight the importance of the selection mechanism. 
When using an oracle selector both \TROVE{} and \PRIMITIVE{} achieve $19\%$ higher accuracy compared to their respective majority-vote baselines. 
This suggests that further gains are more likely to come from better selection rather than increased prompt diversity alone.


\section{Related Work}

\subsection{Library Learning on Formal Syntaxes}
Early work in library learning leverages domain‐specific languages (DSLs) to constrain the search space and extract reusable abstractions.  
DreamCoder alternates between a neural–guided enumerative \textit{wake} phase that finds programs in a functional DSL and a \textit{sleep} phase that compresses recurring subprograms into a growing library of abstractions~\citep{ellis2023dreamcoder}.  
Based on this, STITCH employs a corpus‐driven, top‐down search to discover higher‐level primitives without exhaustive enumeration~\citep{bowers2023top}, while Babble uses e‐graphs and anti‐unification to merge and refine candidate subexpressions into a library~\citep{cao2023babble}.  
LILO further augments DreamCoder by integrating an LLM to propose promising enumerative candidates alongside classic search, improving both scalability and solution quality~\citep{grand2023lilo}.

\subsection{LLM Program Creation on MATH}
More recent approaches drop the DSL constraint and operate directly in general‐purpose languages via in‐context “toolboxes.”  
LATM prompts an LLM once per task to generate exactly one helper function from Big-Bench, then uses that toolbox to solve the task~\citep{cai2023large}.  
CREATOR and CRAFT both adopt multi‐iteration pipelines: CREATOR cycles through generate–refine–execute stages~\citep{qian2023creator}, while CRAFT adds abstraction, validation, deduplication, and retrieval steps to assemble a specialized toolset~\citep{yuan2023craft}.  
ReGAL, which was published parallely to \TROVE{}, applies offline refactoring and pruning cycles to Python solutions on MATH~\citep{stengel2024regal}.  
A recent study suggests that \TROVE{} outperforms CREATOR, LATM, and CRAFT on the MATH benchmark in both accuracy and efficiency~\citep{wang2024tools}.  


\noindent Within automated theorem proving, several works extract reusable formal lemmas from ground-truth proofs. ATG~\citep{lin2024atg} generates new, reusable theorems from the Metamath library to improve downstream proof search, while REFACTOR~\citep{zhou2024refactor} extracts modular lemmas from existing proofs to shorten and strengthen theorem proving. LEGO-Prover~\citep{wang2023lego} introduces a dynamic library of verified lemmas and leverages an LLM to construct proofs modularly, enabling the system to tackle increasingly complex theorems more effectively.

\subsection{Evaluation of Tool Use and Library Learning}
Empirical analyses have begun to question whether toolbox learning truly drives gains over simple sampling.  
Most related to our work, \citet{berlot2024library} demonstrate that for \TROVE{} and LEGO-Prover most ``learned`` libraries are invoked only once, casting doubt on their reusability.
In a concurrent study, they show that for LEGO-Prover the benefit of the toolbox mechanism vanishes, as soon as the computational cost is taken into account~\citep{berlot2025llm}.
Regarding sampling in Program Synthesis, ~\citet{li2024programming} show that finetuning an LLM and repeatedly sampling from it significantly outperforms library learning methods such as LILO~\citep{grand2023lilo} and ReGAL~\citep{stengel2024regal}.
However, in contrast to \TROVE{} these approaches leverage an offline dataset to create the library, whereas \TROVE{} creates its toolbox online during inference.
More broadly, \citet{yue2025does} find that increasing the number of samples from a base LLM often outperforms reinforcement learning methods under a fixed compute budget. This observation aligns with our findings on \TROVE{}, suggesting that, in some cases, simply allocating more compute to sampling from a primitive model can match or exceed the performance of more complex mechanisms such as toolbox construction.


\section{Conclusion}
\label{sec:conclusion}

The findings of this study indicate that the primary advantage of the \TROVE{} approach comes less from the use of an incrementally assembled toolbox of abstractions and more from the increased probability of finding the correct solutions through repeated sampling. 
When controlling for the total number of generated solutions, the baseline approach that simply samples more candidate programs using its inherent domain knowledge performs comparable to \TROVE{}. 
In the MATH domain, we confirm that library learning does not bring a significant advantage over this simple baseline.
Preliminary results indicate that this observation also applies to further domains (TabMWP, WTQ and HiTab).
However, our experiments show that \TROVE's prompting mechanism does lead to an increased hypothesis space.
Under a stronger selection mechanism, this may lead to a small benefit over \PRIMITIVE{}, which exists in our experiments but under majority selection we demonstrate it is not significant.
More importantly, a driving factor is the choice of the selection mechanism. 
Although \TROVE{} and \PRIMITIVE{} often create a correct candidate, the selection mechanism often does not select it, as shown by the $19\%$ difference between oracle selection and majority voting for $K=15$.
Improving the current selection mechanism would lead to a much higher benefit than the one currently brought from \TROVE{} over the baseline. 

In conclusion, while our results cast doubt on whether tool-based methods represent an alternative to intensive sampling in MATH, we remain optimistic about the long-term potential of systematic abstraction learning for LLMs in other problem areas. 
Recent work on agentic tasks~\citep{zhiruo2025inducing,zheng2025skillweaver} has shown that tools can effectively compress past experiences that can then be reused to improve efficiency and solving harder long horizon challenges.
By deepening understanding on when and how library learning can assist LLMs, future research can more effectively utilize toolboxes to achieve improvements in complex problem solving.

\section{Limitations}
A main limitation of our work is that our re-evaluation is primarily focused on the MATH dataset.
Although preliminary studies suggest that the observation also applies to TabMWP, WTQ and HiTab, they need to be further analyzed.

Second, we evaluate both approaches exclusively using the \CODELLAMATWO{} model.
Although this aligns with the original evaluation of \TROVE{}, further experiments involving a broader range of LLMs would help validate the generality of our conclusions.

Third, we assess \TROVE{}’s upper-bound performance under an oracle selection mechanism, which assumes access to the correct answer within the candidate set. While this provides insights into the potential benefits of prompt diversity, it is not applicable in practice. We do not propose or evaluate any realistic selection strategy capable of consistently leveraging this potential, leaving this an interesting direction for future work.


Moreover, our analysis is limited to quantitative performance. We do not perform a qualitative investigation into the nature of the tools generated or their reuse, which lies outside the scope of this work. For a detailed examination of tool reuse in \TROVE{}, we refer the reader to \citet{berlot2024library} and Appendix~\ref{apx:toolreuse}.


Finally, we use the number of LLM calls as a proxy for computational cost when comparing the two approaches. However, this approximation does not imply equal runtime performance. In practice, even with the same number of LLM calls, the \PRIMITIVE{} baseline exhibits faster execution, likely due to more efficient batching and reduced overhead from its simpler structure.

\section*{Impact Statement}
This paper advances Library Learning by analyzing the effectiveness of toolbox-based approaches for program generation with LLMs in mathematical problem solving. Our findings suggest that reported improvements from toolbox reuse may largely stem from increased sampling rather than true abstraction learning. This informs a broader understanding of when abstraction mechanisms in LLMs are beneficial, potentially guiding more efficient prompting and evaluation strategies.
From an ethical perspective, this work does not involve sensitive data, user interaction, or deployment in high-stakes environments. While insights may influence future tool design for educational or productivity applications, we foresee no immediate societal risks.

\section*{Acknowledgements}
This work is partially supported by the Federal Ministry of Education and Research (BMBF) and by the Federal Ministry for Economic Affairs and Energy of Germany (BMWE).

\bibliography{icml2025}
\bibliographystyle{icml2025}

\newpage
\appendix
\onecolumn
\section{Experimental Setup}
\label{apx:exp_setup}
We used the hyperparameters specified in the original \TROVE{} paper~\citep{wang2024trove}, except for the \PRIMITIVE{} baseline that was extended for different values of the num\_return\_sequences parameter, which defines the number of LLM calls per task. 
Furthermore, per MATH domain, the \PRIMITIVE{} baseline and \TROVE{} were run for five different seeds.
A slight adaptation was performed by setting the \textit{exec\_timeout} value from $100s$ to $30s$, but this did not lead to any performance drop.
We used the official implementation from \href{https://github.com/zorazrw/trove}{Github}, which is released under the CC-BY-SA-4.0 license.

The experiments were run on a single NVIDIA RTX A6000, 2 CPUs, and 48GB RAM.
Our code and data are available at \href{https://github.com/Tsesterh/TroVE_Compute_Matched}{https://github.com/Tsesterh/TroVE\_Compute\_Matched}.

As shown in Table~\ref{tab:runtimes}, although we match \PRIMITIVE{} in terms of LLM call budget, the baseline method runs consistently faster than \TROVE{}. We propose that this is due to \PRIMITIVE{} batching 15 LLM calls, whereas \TROVE{} iteratively calls the LLM 5 times per mode and performs additional postprocessing, i.e., periodical toolbox trimming.

\begin{table}[h]
    \centering
    \caption{Hyperparameters for TroVE and the \PRIMITIVE{} baseline.}
    \begin{tabular}{l r r}
        \hline
        \textbf{Hyperparameter} & \textbf{TroVE} & \textbf{\PRIMITIVE{}} \\
        \hline
        trim\_steps & 500 & 500 \\
        exec\_timeout & 100 & 100 \\
        top\_p & 0.95 & 0.95 \\
        num\_return\_sequences & 5 & 1, ..., 15 \\
        temperature & 0.6 & 0.6 \\
        max\_new\_tokens & 512 & 512 \\
        suffix & --- & primitive \\
        \hline
    \end{tabular}
\end{table}

\begin{table}[h!]
\centering
\small
\begin{tabular}{lccccccc|c}
\toprule
\textbf{Method} & \textbf{Algebra} & \textbf{Geometry} & \textbf{Counting} & \textbf{Intermediate} & \textbf{Number} & \textbf{Prealgebra} & \textbf{Precalculus} & \textbf{MATH} \\
\midrule
\PRIMITIVE{} & 25 & 5 & 7 & 14 & 12 & 17 & 4 & 84 \\
\TROVE{}     & 38 & 12 & 12 & 25 & 26 & 28 & 7 & 148 \\
\bottomrule
\end{tabular}
\caption{Runtimes (in hours) per category for \PRIMITIVE{} and \TROVE{}. We report the mean across $5$ seeds rounding to the nearest hour.}
\label{tab:runtimes}
\end{table}

\section{Additional Experimental Results}
\label{apx:additional_experiments}
In this section, we provide additional insights by analyzing \TROVE{} and \PRIMITIVE{} on different topics.
In \S\ref{apx:reproduction} we show that the results of the original work probably were produced by prompting \PRIMITIVE{} only once.
In \S\ref{apx:cov_diversity} we analyze differences in the sets of solutions found by \TROVE{} and \PRIMITIVE{}, evaluating the question whether \TROVE{}'s diverse prompting mechanisms can solve challenges that cannot be solved by \PRIMITIVE{}.
In \S\ref{apx:jaccard} we further analyze the similarity of the proposed solutions between \PRIMITIVE{} and each of \TROVE{}'s modes.
In \S\ref{apx:selection_mechanism} we analyze the potential of each mode for an increased computational budget, combining the results of different seeds to evaluate for $K=75$.
Lastly, in \S\ref{apx:math_distribution} we analyze whether there is a difference between the difficulty levels that both approaches solve on MATH.

\subsection{Reproducing the \TROVE{} paper}
\label{apx:reproduction}
To test the hypothesis that the \PRIMITIVE{} baseline is outperformed by \TROVE{} due to the computational budget, we reproduce the results for different computational budgets $K$.
As can be seen in Table \ref{tab:sampling_orig} for a computational budget of $K=1$ the performance of the \PRIMITIVE{} baseline is closest to the performance reported in the original paper.
Furthermore, when aligned for compute, \PRIMITIVE{} performs equally as \TROVE{}.

\begin{table}[ht]
\centering
\small
\caption{Performance of \PRIMITIVE{} and \TROVE{} for different computational budgets using the agreement-based selection mechanism (for \TROVE{} we report the results for the improved version of the mechanism). We report for multiples of $3$ as for \TROVE{} each of the three modes needs to be called. $P@K$ stands for calling the \PRIMITIVE{} mode $K$ times, $T@K$ for calling \TROVE{} mode $K$ times. As for $K=1$ no \TROVE{} value exists, we add the originally reported value of \PRIMITIVE{}, to show that it was indeed not aligned for compute, as the values are within $2\%$ on the whole MATH dataset. We report the mean values across $5$ seeds.}
\label{tab:sampling_orig}
\begin{tabular}{l rr | rr | rr | rr | rr | rr}
\toprule
Category & P@1 & P (orig) & P@3 & T@3 & P@6 & T@6 & P@9 & T@9 & P@12 & T@12 & P@15 & T@15 \\
\midrule
algebra & 0.13 & 0.15 & 0.19 & \textbf{0.23} & 0.23 & \textbf{0.26} & 0.25 & \textbf{0.28} & 0.26 & \textbf{0.29} & 0.27 & \textbf{0.29} \\
counting & 0.16 & 0.14 & 0.20 & \textbf{0.20} & 0.22 & \textbf{0.24} & 0.22 & \textbf{0.26} & 0.22 & \textbf{0.27} & 0.24 & \textbf{0.27} \\
geometry & 0.05 & 0.06 & \textbf{0.06} & 0.05 & \textbf{0.08} & 0.06 & \textbf{0.08} & 0.07 & \textbf{0.08} & 0.07 & \textbf{0.08} & 0.08 \\
intermediate & 0.07 & 0.05 & 0.10 & \textbf{0.10} & \textbf{0.12} & 0.12 & 0.12 & \textbf{0.13} & 0.13 & \textbf{0.13} & \textbf{0.14} & 0.13 \\
number & 0.20 & 0.16 & \textbf{0.24} & 0.24 & 0.26 & \textbf{0.26} & 0.28 & \textbf{0.28} & 0.28 & \textbf{0.29} & 0.28 & \textbf{0.30} \\
prealgebra & 0.20 & 0.21 & \textbf{0.26} & 0.25 & \textbf{0.30} & 0.29 & \textbf{0.32} & 0.31 & \textbf{0.32} & 0.32 & \textbf{0.33} & 0.32 \\
precalculus & 0.06 & 0.10 & 0.10 & \textbf{0.13} & 0.12 & \textbf{0.17} & 0.14 & \textbf{0.18} & 0.14 & \textbf{0.20} & 0.15 & \textbf{0.20} \\
\midrule
MATH & 0.14 & 0.12 & 0.18 & \textbf{0.19} & 0.21 & \textbf{0.22} & 0.23 & \textbf{0.24} & 0.23 & \textbf{0.25} & 0.24 & \textbf{0.25} \\
\bottomrule
\end{tabular}
\end{table}

\subsection{Coverage Diversity}
\label{apx:cov_diversity}
To quantify the additional coverage the diversity of the three modes brings, we separate the additional benefit of \TROVE{} over \PRIMITIVE{} into two parts, assuming a perfect selection mechanism:

\begin{enumerate}[leftmargin=*, itemsep=2pt]
  \item \textbf{Consistent gain} –
        tasks consistently solved by \emph{every} \TROVE{} seed and by \emph{no}
        \PRIMITIVE{} seed.
  \item \textbf{Potential gain} –
        tasks solved by \emph{at least one} \TROVE{} seed,
        by \emph{no} \PRIMITIVE{} seed,
        but \emph{not} solved by all \TROVE{} seeds. These represent tasks in the hypothesis space that can be reached basically in addition by a \TROVE{} seed but not by a \PRIMITIVE{} seed.
\end{enumerate}

Table \ref{tab:trove-benefit} shows both the numbers per MATH category
and for the benchmark as a whole, as well for \TROVE{} and for \PRIMITIVE{}.
The results show that over the different \TROVE{} seeds the method consistently solves $1.03\%$ of tasks that were not solved by any \PRIMITIVE{} run.
Potentially, on the five seeds, \TROVE{} additionally solves $8.28\%$ of the tasks in any seed that were not found by any \PRIMITIVE{} seed.
On the other hand, \PRIMITIVE{} also solves $0.75\%$ that were not solved by any \TROVE{} run.
Therefore, the results show that the different prompt modes do not necessarily help \TROVE{} solve a larger variety of challenges.

\begin{table*}[t]
      \centering
          \caption{Exclusive task coverage by \TROVE{} and \PRIMITIVE{} across MATH categories. For each method, we report the number and percentage of tasks solved (i) consistently across all five seeds but by no seed of the other method (consistent gain), and (ii) by at least one seed but not all seeds, and not by any seed of the other method (potential gain).}
    \label{tab:trove-benefit}
      \setlength{\tabcolsep}{6pt}
      \small
    \begin{tabular}{lrrrrrrrr}
    \toprule
    & \multicolumn{4}{c}{\textbf{TROVE}} & \multicolumn{4}{c}{\textbf{PRIMITIVE}} \\
    \cmidrule(lr){2-5}\cmidrule(lr){6-9}
    & \multicolumn{2}{c}{Consistent gain} & \multicolumn{2}{c}{Potential gain} & \multicolumn{2}{c}{Consistent gain} & \multicolumn{2}{c}{Potential gain}\\
    \cmidrule(r){2-3}\cmidrule(r){4-5}\cmidrule(r){6-7}\cmidrule(r){8-9}
    Category & \# & \% & \# & \% & \# & \% & \# & \% \\
    \midrule
    Algebra      &  25 &  2.84\% &  83 &  9.42\% &  10 &  1.14\% &  84 &  9.53\% \\
    Counting     &   3 &  1.03\% &  41 & 14.09\% &   5 &  1.72\% &  18 &  6.19\% \\
    Geometry     &   0 &  0.00\% &  18 &  7.59\% &   1 &  0.42\% &  26 & 10.97\% \\
    Intermediate &   2 &  0.40\% &  60 & 11.93\% &   7 &  1.39\% &  32 &  6.36\% \\
    Number       &   1 &  0.20\% &  49 &  9.86\% &   0 &  0.00\% &  37 &  7.44\% \\
    Prealgebra   &   2 &  0.31\% &  64 & 10.06\% &   2 &  0.31\% &  60 &  9.43\% \\
    Precalculus  &   1 &  0.64\% &  18 & 11.54\% &   0 &  0.00\% &  12 &  7.69\% \\
    \midrule
    \textbf{MATH}     &  33 &  1.03\% & 265 &  8.28\% &  24 &  0.75\% & 218 &  6.81\% \\
    \bottomrule
    \end{tabular}

\end{table*}

\subsection{Comparing \PRIMITIVE{} and \TROVE{} solutions}
\label{apx:jaccard}
For the cross–type analysis we compare every \PRIMITIVE{} run with every run of a given \TROVE{} mode \(M\).
Let  

\[
\mathcal{S}_{\text{P}}=\{1,\dots ,k\}, \qquad
\mathcal{S}_{M}       =\{1,\dots ,\ell\}
\]

be the sets of indexes of random seeds for \PRIMITIVE{} and for mode \(M\),
and denote by \(A^{\text{P}}_{s}\) and \(A^{M}_{t}\) the corresponding sets of
solved challenges.  
The \emph{cross–type mean Jaccard similarity} between \PRIMITIVE{} and the
mode \(M\) is then

\[
\overline{J}_{\text{P}\leftrightarrow M}
\;=\;
\frac{1}{k\,\ell}\,
\sum_{s\in\mathcal{S}_{\text{P}}}\;
\sum_{t\in\mathcal{S}_{M}}
\underbrace{\frac{\lvert A^{\text{P}}_{s}\cap A^{M}_{t}\rvert}
                 {\lvert A^{\text{P}}_{s}\cup A^{M}_{t}\rvert}}_{J(A^{\text{P}}_{s},A^{M}_{t})},
\]

i.e., the average Jaccard similarity over all \(k\!\times\!\ell\)
pairs of runs from the two types.  
We compute this value separately for each \TROVE{} mode
(\textsc{Skip}, \textsc{Import}, \textsc{Create}) and for every
problem category.

\begin{figure}[h!]
    \centering
    \includegraphics[width=0.75\linewidth]{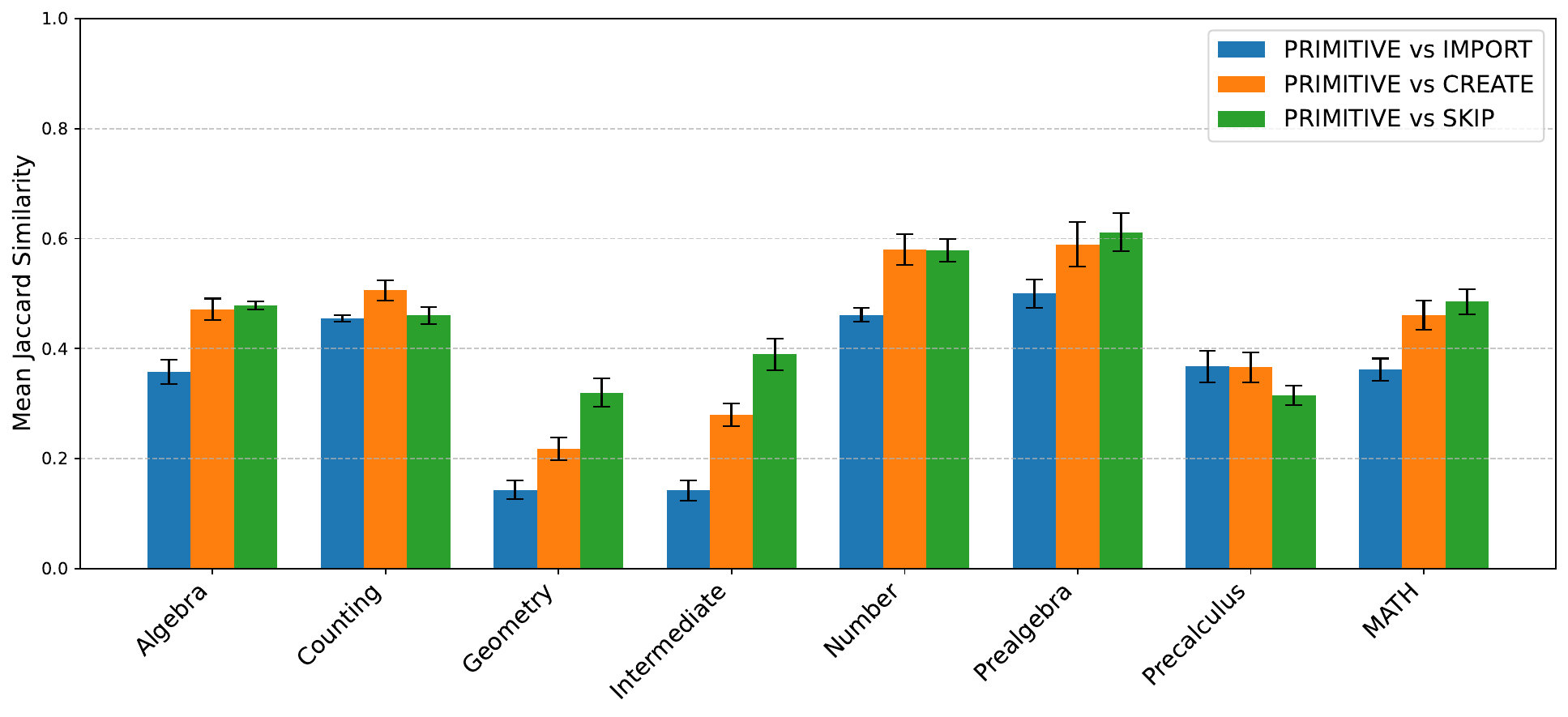}
    \caption{Mean Jaccard Similarities over $5$ random seeds between \PRIMITIVE{} and each \TROVE{} mode for solutions found by the oracle selection mechanism.}
    \label{fig:jaccard_across_modes}
\end{figure}

As depicted in Figure~\ref{fig:jaccard_across_modes}, when computing the Jaccard similarity between \PRIMITIVE{} and the three modes, \textsc{Skip} and \PRIMITIVE{} have the highest value (Figure \ref{fig:jaccard_across_modes}).
This makes sense as \textsc{Skip} and \PRIMITIVE{} share the same prompt.
Furthermore, \PRIMITIVE{} is least similar to \IMPORT{}, as their prompting styles differ the most.
However, this does not count for each category, as for precalculus the similarity to \IMPORT{} is actually highest.
We suggest that the reason for this is that \IMPORT{} outperforms \SKIP{} in precalculus, as shown in Figure~\ref{fig:solved_across_modes}, resulting in a potentially greater overlap with \PRIMITIVE{}.

\subsection{Impact of Selection Mechanism}
\label{apx:selection_mechanism}
We analyze the potential of each approach for an increasing computational budget assuming a perfect oracle selection mechanism.
Table~\ref{tab:oracle_accuracies_15} shows that \TROVE{} slightly outperforms \PRIMITIVE{} in different values of $K$, mostly consistent by $1\%$.

\begin{figure}[h!]
    \centering
    \includegraphics[width=0.55\linewidth]{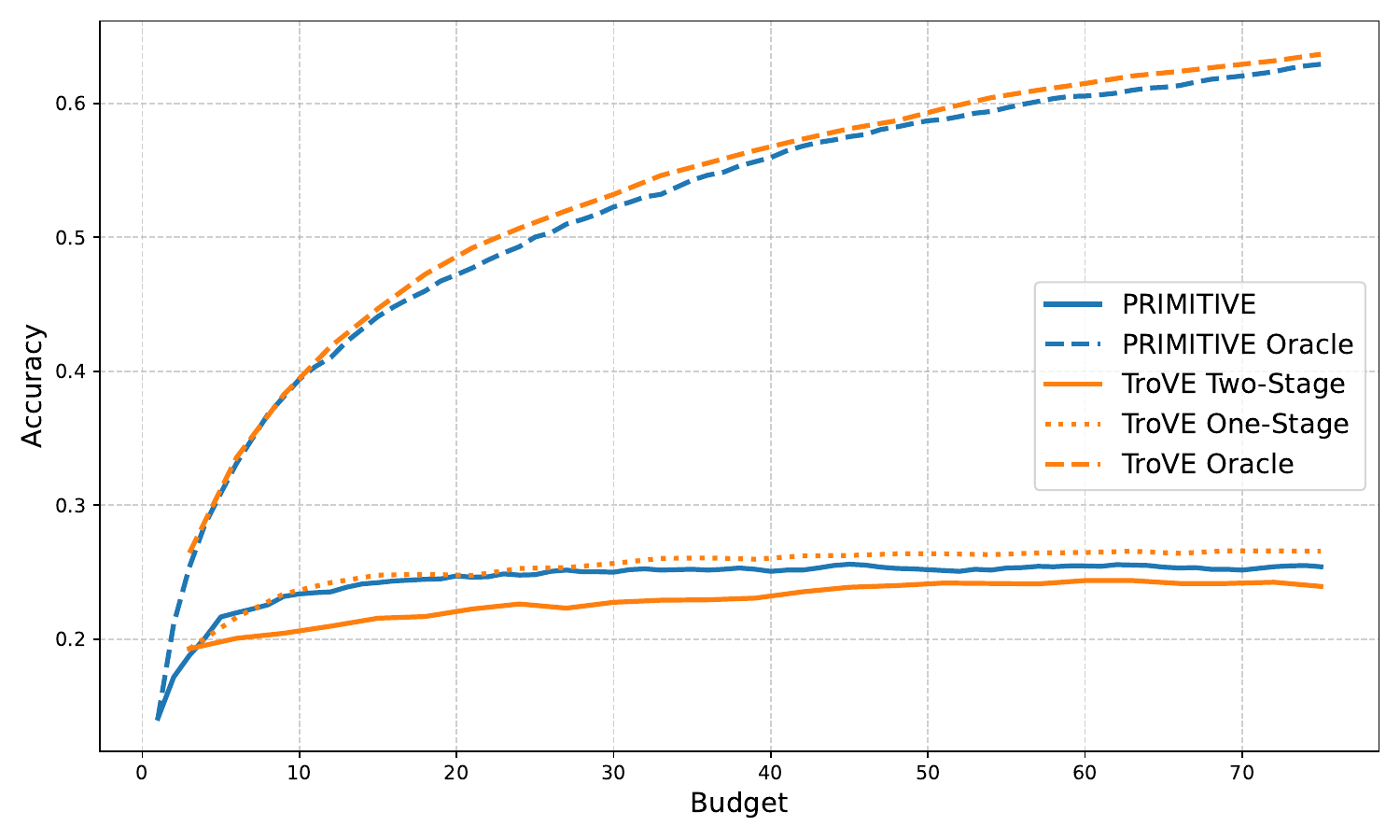}
    \caption{Accuracies for \PRIMITIVE{} and \TROVE{} across different selection mechanisms up to a sample size of $75$. \TROVE{} Two-Stage refers to the original implementation as discussed in Section \ref{sec:correcting_selection}, \TROVE{} One-Stage for the corrected mechanism.}
    \label{fig:n_75_accuracies}
\end{figure}

When combining the maximum $15$ answers of the $5$ seeds, we can plot the performance for up to $75$ samples.
Figure~\ref{fig:n_75_accuracies} depicts how the accuracy changes for each of the approaches presented in this work.
Interestingly, also for the oracle as for the one-stage agreement-based selection mechanism, \TROVE{} slightly outperforms \PRIMITIVE{}, for the oracle by $0.75\%$ after $75$ samples and for the agreement-based method by $2.62\%$.
However, for the agreement-based selection modes, the performance does not anymore increase significantly for \TROVE{} nor for \PRIMITIVE{}.
Therefore, to leverage the full potential of both approaches, it may be worth researching more evolved selection mechanisms.

\begin{table}[ht]
\centering
\caption{Primitive (P) and Trove (T) $pass@K$ values (using the oracle selection mechanism) for different values of $K$. We report the mean value across $5$ random seeds.}
\label{tab:oracle_accuracies_15}
\small
\begin{tabular}{l rr | rr | rr | rr | rr | rr}
\toprule
Category & P@1 & T@1 & P@3 & T@3 & P@6 & T@6 & P@9 & T@9 & P@12 & T@12 & P@15 & T@15 \\
\midrule
Algebra & 0.13 & \,--- & 0.24 & \textbf{0.29} & 0.32 & \textbf{0.36} & 0.38 & \textbf{0.41} & 0.41 & \textbf{0.44} & 0.44 & \textbf{0.47} \\
Counting & 0.16 & \,--- & 0.27 & \textbf{0.28} & 0.34 & \textbf{0.35} & 0.38 & \textbf{0.40} & 0.42 & \textbf{0.43} & 0.44 & \textbf{0.46} \\
Geometry & 0.05 & \,--- & \textbf{0.11} & 0.09 & \textbf{0.18} & 0.16 & \textbf{0.22} & 0.19 & \textbf{0.25} & 0.22 & \textbf{0.28} & 0.24 \\
Intermediate & 0.07 & \,--- & \textbf{0.14} & 0.12 & \textbf{0.19} & 0.18 & \textbf{0.23} & 0.22 & \textbf{0.25} & 0.25 & \textbf{0.28} & 0.27 \\
Number & 0.20 & \,--- & \textbf{0.31} & 0.31 & \textbf{0.40} & 0.39 & 0.44 & \textbf{0.45} & 0.47 & \textbf{0.48} & 0.50 & \textbf{0.52} \\
Prealgebra & 0.20 & \,--- & 0.35 & \textbf{0.36} & 0.44 & \textbf{0.45} & 0.49 & \textbf{0.50} & 0.53 & \textbf{0.54} & 0.56 & \textbf{0.57} \\
Precalculus & 0.06 & \,--- & 0.12 & \textbf{0.17} & 0.19 & \textbf{0.23} & 0.22 & \textbf{0.26} & 0.26 & \textbf{0.29} & 0.29 & \textbf{0.31} \\
\midrule
MATH & 0.14 & \,--- & 0.24 & \textbf{0.26} & 0.32 & \textbf{0.33} & 0.37 & \textbf{0.38} & 0.40 & \textbf{0.41} & 0.43 & \textbf{0.44} \\
\bottomrule
\end{tabular}

\end{table}

\subsection{Distribution among MATH difficulties}
\label{apx:math_distribution}

Each task in MATH is further labeled with a difficulty level from 1 (easiest) to 5 (most difficult), allowing for fine-grained analysis of model performance across both topic and complexity. Table~\ref{tab:math-breakdown} provides a breakdown of the dataset by category and difficulty level.

\begin{table}[h]
  \centering
  \caption{Distribution of problems in the MATH dataset by category and difficulty level (L1--L5).}
  \label{tab:math-breakdown}
\begin{tabular}{lrrrrrr}
\toprule
Category & Size & Level 1 & Level 2 & Level 3 & Level 4 & Level 5 \\
\midrule
Algebra & 881 & 125 & 159 & 192 & 209 & 196 \\
Counting & 291 & 31 & 77 & 59 & 69 & 55 \\
Geometry & 237 & 29 & 45 & 50 & 63 & 50 \\
Intermediate & 503 & 28 & 68 & 116 & 146 & 145 \\
Number & 497 & 29 & 81 & 112 & 132 & 143 \\
Prealgebra & 636 & 61 & 149 & 178 & 124 & 124 \\
Precalculus & 156 & 29 & 29 & 34 & 28 & 36 \\
\bottomrule
\end{tabular}
\end{table}

We analyze whether \TROVE{} differs from \PRIMITIVE{} in the distribution of solved difficulties.
For this, the oracle mechanism is chosen again to evaluate the potential of both \TROVE{} and \PRIMITIVE{}.
Table \ref{tab:levels} shows that both approaches solve a higher proportion of simple tasks than difficult ones.
While they find a solution in more than $70\%$ of the tasks of level 1, at level 5 they only solve roughly $20\%$. 
Although \TROVE{} solves slightly more tasks at difficulty level 5, again there is no significant difference between both approaches.
This suggests that \TROVE{}'s prompt diversity does not positively influence the performance on harder tasks.

\begin{table*}[h!]
\centering
\small
\caption{Analysis of the difficulty levels solved by \TROVE{} and \PRIMITIVE{} (in \%) for $pass@K$ with $K = 15$ on the MATH dataset. The mean value across $5$ random seeds is reported.}
\label{tab:levels}
\begin{tabular}{llrrrrr}
\toprule
 &  & Level 1 & Level 2 & Level 3 & Level 4 & Level 5 \\
Category & Method &  &  &  &  &  \\
\midrule
\multirow{2}{*}{Algebra} & Primitive & 75.52 & 59.37 & 47.92 & 31.96 & 19.18 \\
 & Trove & 84.00 & 65.28 & 52.08 & 33.40 & 18.78 \\
\cline{1-7}
\multirow{2}{*}{Counting} & Primitive & 76.77 & 63.12 & 46.78 & 31.30 & 13.45 \\
 & Trove & 79.35 & 63.64 & 52.88 & 30.14 & 15.64 \\
\cline{1-7}
\multirow{2}{*}{Geometry} & Primitive & 38.62 & 44.44 & 25.20 & 27.30 & 10.80 \\
 & Trove & 35.86 & 38.22 & 24.00 & 22.54 & 7.60 \\
\cline{1-7}
\multirow{2}{*}{Intermediate} & Primitive & 60.00 & 46.47 & 35.00 & 23.01 & 13.79 \\
 & Trove & 64.29 & 48.82 & 35.00 & 20.68 & 10.21 \\
\cline{1-7}
\multirow{2}{*}{Number} & Primitive & 85.52 & 68.15 & 57.50 & 45.45 & 30.35 \\
 & Trove & 83.45 & 66.91 & 57.14 & 47.12 & 36.50 \\
\cline{1-7}
\multirow{2}{*}{Prealgebra} & Primitive & 79.02 & 71.68 & 55.17 & 51.29 & 31.94 \\
 & Trove & 80.98 & 70.47 & 58.31 & 52.90 & 32.74 \\
\cline{1-7}
\multirow{2}{*}{Precalculus} & Primitive & 64.14 & 41.38 & 18.82 & 19.29 & 10.00 \\
 & Trove & 71.72 & 33.79 & 20.00 & 29.29 & 8.33 \\
\cline{1-7}
\multirow{2}{*}{Math} & Primitive & 71.63 & 60.63 & 46.13 & 34.79 & 20.96 \\
 & Trove & 76.02 & 61.22 & 48.37 & 35.15 & 21.34 \\
\cline{1-7}
\bottomrule
\end{tabular}
\end{table*}

\subsection{Tool Reuse}
\label{apx:toolreuse}
As this work complements the work by \citet{berlot2024library}, it focuses solely on the quantitative re-evaluation of \TROVE{}.
However, \citet{berlot2024library} perform a deeper analysis on tool reuse on \textsc{MATH}, across all seven categories. Their findings show that \TROVE{} only learns functions for 3 of the 7 areas --- counting, number, and prealgebra --- resulting in a total of just 15 learned functions. No functions are learned for algebra, geometry, intermediate algebra, or pre-calculus. Moreover, only two of these functions are ever reused in a correct solution: \texttt{is\_perfect\_square(n)} once and \texttt{is\_prime(num)} twice. With just three reuses across 3,201 test questions, the authors conclude that \TROVE{}’s performance gains are unlikely to stem from tool reuse, which is elaborated in our work by showing that the benefit simply comes from increased compute.

\subsection{Analysis on Other Domains} \label{apx:domains}

To extend our evaluation to additional domains, we test the \PRIMITIVE{} baseline on three more datasets: TabMWP \citep{lu2022dynamic}, WTQ \citep{pasupat2015compositional}, and HiTab \citep{cheng2021hitab}.
We run experiments using a single random seed and three different computational budgets ($K \in {1, 5, 15}$).
As shown in Figure~\ref{tab:datasets_comparison}, when compute-matched with $K = 15$, \PRIMITIVE{} performs much more similarly to \TROVE{} than suggested by the originally reported results.
Specifically, accuracy improves by 3\% on TabMWP, 5\% on WTQ, and 6\% on HiTab.


To validate these trends more robustly, we propose evaluating across multiple random seeds in future work.
Nonetheless, these preliminary results suggest that our core observations from \textsc{MATH} may generalize to other domains as well, indicating that the performance gap between \TROVE{} and \PRIMITIVE{} may be marginal when computational budgets are matched.

\begin{table}[ht]
\centering
\begin{tabular}{lccc}
\toprule
\textbf{Method} & \textbf{TabMWP} & \textbf{WTQ} & \textbf{HiTab} \\
\midrule
Primitive $K=1$   & 0.36 & 0.18 & 0.08 \\
Primitive $K=5$   & 0.42 & 0.23 & 0.14 \\
Primitive $K=15$  & \textbf{0.46} & \textbf{0.25} & \textbf{0.15} \\
\midrule
Primitive \citep{wang2024trove} & 0.43 & 0.20 & 0.09 \\
\TROVE{} \citep{wang2024trove}     & 0.47 & 0.21 & 0.18 \\
\bottomrule
\end{tabular}
\caption{Accuracy across further datasets. Results show that with matched compute ($K=15$), \PRIMITIVE{} approaches the performance of \TROVE{}.}
\label{tab:datasets_comparison}
\end{table}

\end{document}